# Field-induced magnetic phase transitions and the resultant giant anomalous Hall effect in antiferromagnetic half-Heusler DyPtBi

H. Zhang[1], Y.L. Zhu[2], Y. Qiu[3], W. Tian[4], H. B. Cao[4], Z. Q. Mao[2], and X. Ke[1]

[1]*Department of Physics and Astronomy, Michigan State University, East Lansing, MI 48824, USA*

[2]*Department of Physics, Pennsylvania State University, University Park, Pennsylvania 16802, USA*

[3]*NIST Center for Neutron Research, National Institute of Standards and Technology, Gaithersburg, Maryland 20899, USA*

[4]*Neutron scattering Division, Oak Ridge National Laboratory, Oak Ridge, Tennessee 37831, USA*

We report field-induced magnetic phase transitions and transport properties of antiferromagnetic DyPtBi. We show that DyPtBi hosts a delicate balance between two different magnetic ground states, which can be controlled by a moderate magnetic field. Furthermore, it exhibits giant anomalous Hall effect ($\sigma^A = 1540\ \Omega^{-1}\ cm^{-1}, \theta_{AHE} = 24\ \%$) in a field-induced Type-I spin structure, presumably attributed to the enhanced Berry curvature associated with avoided band-crossings near the Fermi energy and / or non-zero spin chirality. The latter mechanism points DyPtBi towards a rare potential realization of anomalous Hall effect in an antiferromagnet with face-center-cubic lattice that was proposed in [Physical Review Letters **87**, 116801 (2001)].

Since the discovery of topological insulators (TIs), integration of non-trivial band topology and magnetism has been long sought after due to both fundamental scientific interest and potential technological applications [1]. The interplay between topology and magnetism is anticipated to give rise to a rich variety of novel topological quantum phenomena. For instance, anomalous quantum Hall effect [2] and axion insulators [3] have been predicted when integrating magnetism with TIs. Another excellent example is the prediction and discovery of Weyl fermions in condensed matter physics that can be induced by the broken time-reversal symmetry [4]. Experimentally, a few magnetic topological materials have been confirmed, such as the intrinsic antiferromagnetic topological insulator $MnBi_2Te_4$ [5], the ferromagnetic Weyl semimetals $Co_3Sn_2S_2$ [6-8], $Co_2MnGa$ [9]. The search for magnetic topological materials exhibiting exotic quantum phenomena has been continuously attracting intense research efforts from the condensed matter physics and materials science communities.

Magnetic metals featuring heavy elements provide an appealing platform to integrate magnetism with the non-trivial band topology. In particular, the rare earth half-Heusler family (LnPtBi) has become a focal point for magnetic topological material research owing to the following merits. First, many members of the LnPtBi family are known to develop long-range magnetic order [10-13]. Second, all elements in LnPtBi are in the fifth row of the periodic table, thus possessing strong spin-orbit coupling (SOC). Third, as a result of the strong SOC, non-trivial band topology is anticipated for LnPtBi due to the critical band touching at Γ point, a feature similar to that of HgTe [14]. For instance, GdPtBi has been predicted to be a candidate of antiferromagnetic TI [3] and large anomalous Hall effect (AHE) has been observed which was attributed to Berry curvature associated with the magnetic field-induced avoided band crossing or Weyl nodes near the Fermi level [11].

In this article, we report magnetic susceptibility, transport, and neutron diffraction studies of DyPtBi, a member of the LnPtBi family. We show that below the Neel temperature $T_N$ = 3.5 K DyPtBi undergoes two field-induced phase transitions at $\mu_0 H_{c1}$ = 1.4 T and $\mu_0 H_{c2}$ = 3.7 T, with the spin structure varying from Type-II to canted Type-I and then to canted Type-II. Interestingly, we observe giant anomalous Hall effect ($\sigma^A = 1540 \ \Omega^{-1} \ cm^{-1}, \theta_{AHE} = 24 \ \%$) concomitant with the canted Type-I antiferromagnetic structure. These phenomena are drastically distinct from those reported in other LnPtBi compounds. This study points towards the intertwined nature of magnetism, electronic band structure and its topology in DyPtBi.

Single crystals of DyPtBi were grown using flux method [15,16]. Magnetic susceptibility measurements of DyPtBi were carried out using a Superconducting Quantum Interference Device (SQUID), and electronic transport measurements were conducted using a Physical Property Measurement System (PPMS). Field dependent single crystal neutron diffraction experiments were performed at High Flux Isotope Reactor (HFIR) in Oak Ridge National Laboratory (ORNL) and NIST Center for Neutron Research (NCNR). The sample was oriented in the (H H L) scattering plane and magnetic field applied along the [1 $\bar{1}$ 0] direction, where H and L are in reciprocal lattice units (*r. l. u.*). Measurements at HFIR were done using the Fixed-Incident Energy Triple-Axis Spectrometer (FIE-TAX) on the HB-1A beam line with a neutron wavelength λ = 2.363 Å and a collimator setting of 40'-40'-40'-80', and the sample was loaded in a vertical field cryomagnet. The experiment at NCNR was carried out on the Multi Axis Crystal Spectrometer (MACS) [17] with a wavelength λ = 2.462 Å and the sample was loaded in a dilution fridge with built-in superconducting magnet. In addition, zero-field single crystal neutron diffraction measurements were also conducted using four-circle neutron diffractometer (HB3A) with a neutron wavelength λ = 1.003 Å at HFIR, ORNL [18].

DyPtBi crystalizes in the space group F-43m (No. 216) with the lattice parameters $a = b = c = 6.6440$ Å and $\alpha = \beta = \gamma = 90°$. Figure 1(a) shows the crystal structure of DyPtBi, in which each constituent element forms a face-center-cubic (FCC) lattice that interpenetrates each other. Temperature dependent magnetic susceptibility $\chi(T)$ measured at $H = 100$ Oe and zero-field longitudinal resistivity $\rho_{xx}(T)$ are shown in Fig. 1(b), and their expanded views in the low temperature region are presented in Fig. 1(c). A sudden drop in $\chi(T)$ is observed below $T_N = 3.5$ K, indicating a paramagnetic to antiferromagnetic phase transition. Concurrently, a sudden increase of $\rho_{xx}(T)$ is observed below $T_N$, suggesting coupling between localized magnetic moments and itinerant electrons. And a semiconducting feature is clearly seen at high temperature, as evidenced by a moderate increase followed by decrease in $\rho_{xx}(T)$. Similar features have been observed in GdPtBi [11]. The slightly smaller value of $T_N$ obtained in $\rho_{xx}(T)$ compared to that in $\chi(T)$ is presumably because $\rho_{xx}(T)$ was measured during cooling down instead of warming up.

The blue curve in Fig. 1(d) shows the isothermal magnetization $M(H)$ measurements at $T = 2$ K with the magnetic field applied along the [0 0 1] direction, which shows several distinct features. There are two sharp increases in the measured magnetization, one at $\mu_0 H_{c1} = 1.4$ T and the other at $\mu_0 H_{c2} = 3.7$ T. As to be discussed next, these two sudden changes in magnetization does not correspond to the more commonly observed spin flop transition [19]; instead, they are associated with phase transitions from one canted antiferromagnetic state to another. As such, we designate three regions with different magnetic structures in [Fig. 2(a)] by region I (0 - $H_{c1}$), region II ($H_{c1}$ - $H_{c2}$) and region III (> $H_{c2}$).

To understand the magnetic structure associated with different regions, we performed single crystal neutron diffraction measurements in the presence of magnetic fields $(\vec{B} \parallel [1\ \bar{1}\ 0])$. Fig. 2(a) shows the neutron diffraction contour map in the [H H L] scattering plane at $T = 0.12$ K and $\mu_0H = 0$ T after background subtraction (background data measured at $T = 7$ K). There are a couple of interesting features which should be pointed out. *First*, nuclear Bragg peaks (for instance, the (0 0 2) Bragg peak) show noticeable residual intensity even after subtracting the background, indicating an enhancement of nuclear Bragg peaks at low temperature. As discussed in the Supplemental Materials [20], the residual nuclear Bragg peak intensity does not indicate existence of ferromagnetic moment, which is in agreement with the $M(H)$ data shown in Fig. 1(d). Instead, it originates from the decrease of extinction of neutron scattering due to the structural distortion which accompanies with the magnetic phase transition [20]. *Second*, multiple magnetic Bragg peaks positioning at (H H L) with half integer values of H and L are clearly observed (for instance, (-5/2 -5/2 L) magnetic peaks are highlighted along the dashed line in Fig. 2(a)), suggesting a propagation wave vector of $\vec{k}_2 = $ (1/2 1/2 1/2). By collecting the magnetic Bragg peak intensities and performing Rietveld refinement [21], the obtained antiferromagnetic structure is illustrated in Fig. 2(d). Dysprosium spins order ferromagnetically within the (1 1 1) plane while spins of neighboring planes along the [1 1 1] direction align antiparallel to each other. Following the conventions used in Ref. [22] for an antiferromagnetic FCC lattice, this spin structure is denoted as Type-II, the same as the magnetic ground state of GdPtBi [11] and YbPtBi [23]. The black curve plotted in Fig. 1(c) represents the temperature dependence of scattering intensity of the (1/2 1/2 1/2) magnetic Bragg peak, which emerges below $T_N$, consistent with the magnetic susceptibility measurement. Detailed information regarding magnetic structure refinement can be found in sections C&D in the Supplemental Materials [20].

Figure 2(b, c) present the background-subtracted neutron diffraction contour map measured at $\mu_0H = 2$ T and 4 T, respectively. Interestingly, at $\mu_0H = 2$ T the magnetic Bragg peaks with half integer values of H and L disappear; Instead, a new set of magnetic Bragg peaks emerge,

which are characterized by a propagation wave vector $\vec{k}_1 = (0\ 0\ 1)$ (also seen in Fig. S1). The corresponding refined magnetic structure is illustrated in Fig. 2(e), where spins are ferromagnetically aligned within the ab-plane, while antiferromagnetically aligned between neighboring ab-planes along the c-axis. This spin configuration is referred to as Type-I [22], which has been observed in NdPtBi [10] and CePtBi [24] measured at zero field. In addition, the residual intensity of nuclear Bragg peaks observed at zero field nearly diminishes at $\mu_0 H = 2$ T, which implies that the lattice distortion lessens and consequently the neutron extinction effects gets enhanced (Fig. S2 and the discussion in section B of the Supplemental Materials). Surprisingly, at $\mu_0 H = 4$ T magnetic Bragg peaks characterized by $\vec{k}_1 = (0\ 0\ 1)$ propagation vector disappear, while magnetic Bragg peaks characterized by $\vec{k}_2 = (1/2\ 1/2\ 1/2)$ re-emerge with weaker intensity, as shown in Fig. 2(c). That is, at $\mu_0 H = 4$ T DyPtBi exhibits Type-II spin structure again. The refined spin structure is presented in Fig. 2(f), where spins point along the c-axis in contrast to the ab-plane spin configuration measured at zero field shown in Fig. 2(d). In addition, a large ferromagnetic component is observed, which is consistent with the $M(H)$ measurement.

The red and black curves in Fig. 1(d) represent the field dependence of neutron scattering intensity of (1/2 1/2 1/2) and (0 0 1) respectively, measured at $T = 1.5$ K. One can see that two transitions between Type-I and Type-II magnetic phases occur at $\mu_0 H_{c1} = 1.4$ T and $\mu_0 H_{c2} = 3.7$ T, which coincides with the sharp jumps in $M(H)$ curves. It was shown that for a perfect antiferromagnetic FCC lattice with only nearest-neighbor interactions, Type-I and Type II spin configurations are degenerate [22]. Our observations in the neutron diffraction measurements suggest that these two spin configurations are nearly degenerate in DyPtBi and that the systems show strong spin-lattice coupling. Slight lattice distortion or a moderate magnetic field can readily tip the delicate balance between these two spin configurations. It is worth noting that such a feature has not been observed in other LnPtBi, such as GdPtBi [11] and TbPtBi [12].

Considering the large SOC and spin-charge coupling in DyPtBi, a natural question arises: what is the electronic response in the presence of magnetic field and how does it correlate to the magnetic structure change? To address this, we first present the longitudinal resistivity ($\rho_{xx}$) measured as a function of magnetic field in Fig. 3 (a). Above $T_N$, $\rho_{xx}(H)$ increases monotonically with increasing field. In contrast, $\rho_{xx}(H)$ shows a much more complex behavior below $T_N$. In region I, $\rho_{xx}(H)$ initially increases slowly and then drops sharply as external field approaches $H_{c1}$; in region II, $\rho_{xx}(H)$ increases slowly before it sharply increases near $H_{c2}$; finally in region III, $\rho_{xx}(H)$ gradually increases with magnetic field. Such highly non-monotonic feature of $\rho_{xx}(H)$ below $T_N$ in DyPtBi is beyond the scope of conventional magnetoresistance theories. Instead, the fact that the sharp changes of $\rho_{xx}$ only occur at $H_{c1}$ and $H_{c2}$ coinciding with the magnetic structure changes signals a significant modification of its electronic structure by the underlying magnetic structure.

The magnetic structure in metallic systems can influence not only electronic band structure, but also band topology. This is exemplified by the modification of the electronic band structure in GdPtBi [11], TbPtBi [12] and CePtBi [24], where the field-induced canted antiferromagnetic structure gives rise to avoided band crossing or Weyl nodes. The resultant enhanced Berry curvature leads to large AHE. Previous angle-resolved photoemission spectroscopy (ARPES) measurements have found that GdPtBi and DyPtBi exhibit similar Fermi surface in the paramagnetic state [25]. Considering that the zero-field magnetic ground state of both compounds has Type-II spin structure, it is reasonable to postulate that the electronic structures of both compounds are similar as well below $T_N$. As non-trivial band topology is predicted for the rare earth half Heusler compounds LnPtBi [14], it is thus tempting to examine the AHE in DyPtBi.

Figure 3(b) shows the Hall resistivity $\rho_{xy}$ measured as a function of field at various temperatures. One can see that $\rho_{xy}(H)$ measured below $T_N$ behaves quite differently from that measured above $T_N$ with additional "bumps" appearing at low fields. In general, $\rho_{xy}$ can be expressed as $\rho_{xy} = \rho_N + \Delta\rho_{xy}$ for magnetic materials. $\rho_N$ refers to the normal Hall effect ($\rho_N \propto H$), and $\Delta\rho_{xy}$ refers to AHE which includes the extrinsic terms arising from skew scattering, side-jump, and the intrinsic terms associated with the spin texture, the Berry curvature due to non-trivial band topology, etc [26]. Since $\rho_{xy}$ at $T = 20$ K is linearly proportional to $H$ [Fig. 3(b)], suggesting that only the normal Hall effect is present at this temperature, we set $\rho_N = \rho_{xy}(20\ K)$. In Fig. 3(c) we plot $\Delta\rho_{xy} = \rho_{xy} - \rho_{xy}(20\ K)$ for measurements done above $T_N$. At each temperature $\Delta\rho_{xy}$ is composed of two over-lapping broad peaks, which decrease in amplitude and move to higher field as the temperature increases. Similar AHE features have been observed in CePtBi [24], GdPtBi [11], and TbPtBi [12], which are mainly attributed to the enhanced Berry curvature in momentum space due to the modified band topology.

Compared to other LnPtBi compounds, the most striking feature of AHE in DyPtBi is the $\Delta\rho_{xy}$ measured below $T_N$ as shown in the lower panel of Fig. 3(d). Three domes in the whole measurement range are clearly observed, corresponding to three different magnetic phases as discussed above [Fig. 2(d-f)]. In the large Hall angle limit, Hall conductivity $\sigma_{yx} = \frac{\rho_{xy}}{(\rho_{xx}^2 + \rho_{xy}^2)}$ is the appropriate measure to quantify the AHE response [26]. In the upper panel of Fig. 3(d) we plot the Hall conductivity measured below $T_N$. $\sigma_{yx}$ measured at $T = 20$ K is also plotted as a reference background. Interestingly, below $T_N$ large $\sigma_{yx}$ shows up only in region II between $\mu_0 H_{c1}$ and $\mu_0 H_{c2}$.

Particularly, $\sigma_{yx}$ exhibits a sharp and narrow peak at $T = 2$ K near $\mu_0 H_{c1}$. This peak becomes broader as temperature increases.

In order to quantify the anomalous Hall response, we turn to the anomalous Hall conductivity $\left(\sigma^A = \sigma_{yx} - \sigma_N\right)$ and anomalous Hall angle $\left[\theta_{AHE} = \frac{\sigma_{yx}}{\sigma_{xx}} - \frac{\sigma_{yx}}{\sigma_{xx}}|_{T=20\,K}\right]$. The normal Hall conductivity $\sigma_N$ is estimated using the Hall data at $20\,K\,(\gg T_N = 3.5\,K)$, i. e. $\sigma_N = \sigma_{yx}\,(20\,K)$. In Fig. 4(a), we show a map of $\sigma^A$ as a function of temperature and magnetic field. We can clearly see a moderate shoulder signal (~ 600 $\Omega^{-1}\,cm^{-1}$) which extends to around 9 K. We also observe a sharp peak in the region of interests (ROI), as labeled by the black frame in Fig. 4(a). The ROI refers to the region with $2\,K < T < T_N = 3.6\,K$ and $\mu_0 H_{c1} < \mu_0 H < \mu_0 H_{c2}$. $\sigma^A$ maximizes at 1540 $\Omega^{-1}\,cm^{-1}$, which is exceptionally large compared to most materials reported thus far [26]. Anomalous Hall angle $\theta_{AHE}$ measured at different temperatures are presented in Fig. 4(b). We find that $\theta_{AHE}$ varies with the magnetic transition. Above $T_N$, a broad peak in $\theta_{AHE}$ is observed, which decreases in amplitude and centers at higher field as temperature increases, a feature similar to GdPtBi [11] and TbPtBi [12]. Below $T_N$, we observe a sharp peak of $\theta_{AHE}$, with the maximum reaching ~ 25% near $\mu_0 H_{c1}$; $\theta_{AHE}$ decreases sharply as the field approaches $\mu_0 H_{c2}$.

To highlight the large magnitude of the observed anomalous Hall response, a comparison of $\sigma^A$ and $\theta_{AHE}$ between DyPtBi and other materials is made in Fig. 4(c). A wide range of $\sigma^A$ has been reported in various materials [6,11,27-32] and different mechanisms of AHE have been proposed, including spin chirality due to non-coplanar spin structure (red), spin texture (green), Berry curvature in momentum space (blue) [6,11,26]. Nevertheless, large $\theta_{AHE}$ (>10%) reported thus far has been exclusive to materials with large Berry curvature in the momentum space. For instance, the large $\sigma^A$ and $\theta_{AHE}$ in GdPtBi [11] and TbPtBi [12] measured below and above $T_N$ have been attributed to the enhanced Berry curvature due to the field-induced Weyl nodes or avoided band crossing of $\Gamma_6 - \Gamma_8$ bands near the Fermi level. We anticipate similar mechanism to be responsible for the AHE observed in DyPtBi above $T_N$, considering similar Fermi surface in the paramagnetic state in GdPtBi and DyPtBi as revealed in ARPES measurements [25]. However, below $T_N$ the anomalous Hall signal in DyPtBi is more intriguing. As described above and shown in Fig. 4(b), in contrast to a broad peak in $\theta_{AHE}$ observed above $T_N$, below $T_N$ a sharp peak of $\theta_{AHE}$ is observed around $\mu_0 H_{c1}$ and $\theta_{AHE}$ decreases sharply to near zero at $\mu_0 H_{c2}$. Furthermore, within region II, both $\sigma^A$ and $\theta_{AHE}$ are significantly enhanced with their maximal values reaching 1540 $\Omega^{-1}\ cm^{-1}$ and 24% respectively, as shown in Fig. 4(a, b, c).

There are a couple of mechanisms that may account for the distinct features observed in region II below $T_N$ in DyPtBi. One is attributed to the avoided band-crossing. It is likely that the altered electronic structure in region II with Type-I spin configuration gives rise to avoided band-crossing points that are even closer to Fermi energy compared to the case above $T_N$ or to the cases in other LnPtBi compounds, which consequently results in enhanced Berry curvature and thus larger AHE. Nevertheless, as shown in Fig. 4(b), we notice that $\theta_{AHE}$ quickly decreases in region II once $H > H_{c1}$, a feature distinct from a broad peak spanning several Telsa observed in other LnPtBi compounds [11,12]. Therefore, although we cannot exclude the effect of avoided band-crossing, we speculate that other mechanism should play a role as well, if not dominates.

Spin chirality mechanism proposed by Shindou and Nagaosa [33] nearly two decades ago offers an alternative and perhaps more probable explanation. This theory predicted that the non-zero spin chirality for Type-I antiferromagnet on a distorted FCC lattice can lead to large anomalous Hall conductivity [33], although the AHE due to spin chirality reported thus far in other material systems is generally small [26]. Our rationales are as follows. First, canted Type-I antiferromagnetic structure is observed exclusively in region II below $T_N$ in DyPtBi, where the anomalous Hall signal ($\sigma^A, \theta_{AHE}$) is the most pronounced. Second, lattice distortion is observed in DyPtBi in region I with Type-II spin structure at low field. As the magnetic field approaches $H_{c1}$, DyPtBi is on the verge of transitioning into type-I AFM structure. Near this phase transition, spin chirality is anticipated to be the most prominent [34], leading to a sharp increase of the anomalous Hall signal. Above $H_{c1}$, the lattice distortion reduces, and the canted ferromagnetic component gets enhanced while the ordered antiferromagnetic moment decreases with increasing magnetic field, which results in decrease of anomalous Hall signal. Finally, when the magnetic field approaches $H_{c2}$, the magnetic structure returns to canted Type-II AFM, leading to the sharp decrease of anomalous Hall signal as observed in Fig. 4(b). Future theoretical studies on the electronic structure, band topology, and their effects on AHE in regions with different magnetic configurations in DyPtBi are warranted.

In summary, we have established DyPtBi as an interesting magnetic topological material that exhibits intricate interplay between magnetism, electronic band structure and band topology. Although its physical behavior above $T_N$ is similar to other LnPtBi cousins, DyPtBi stands out as a system displaying sequential field-induced magnetic phase transitions, which suggests delicate energetic balance between different magnetic ground states. Giant anomalous Hall signal ($\sigma^A = 1540 \: \Omega^{-1} \: cm^{-1}, \theta_{AHE} = 24 \:\%$) is observed in the intermediate field region, accompanied by the emergence of Type-I antiferromagnetic structure. This feature is ascribed to the enhanced Berry curvature associated with avoided band-crossings near the Fermi energy and / or non-zero spin chirality, the latter of which renders this system a rare experimental realization of AHE in antiferromagnets on a distorted FCC lattice.

**Figure Captions**

**Figure** 1 (a) Crystal structure of DyPtBi. Dysprosium atoms are in blue, platinum atom in silver and bismuth atom in purple. (b) Magnetic susceptibility (blue) measured at an applied magnetic field of 100 Oe and longitudinal resistivity (red) of DyPtBi measured as a function of temperature. (c) An expanded view of (b) in the low temperature region, which is over-plotted with the order parameter scan of magnetic Bragg peak (0.5 0.5 0.5). (d) Isothermal magnetization $M(\mu_0 H)$ curve (blue) of DyPtBi measured at $T = 2$ K (upper panel), which is over-plotted with the field dependence of neutron diffraction intensity of two characteristic magnetic Bragg peak (0 0 1) and (0.5 0.5 0.5) measured at $T = 1.5$ K.

**Figure** 2 (a-c) Neutron diffraction intensity contour map in the [H H L] scattering plane measured in the presence of different magnetic fields applied along the [1$\bar{1}$0] direction. The data measured at $T = 7$ K were used for background subtraction. The powder rings come from the small residual Aluminum signal from both sample can and sample holder. (d-f) The corresponding antiferromagnetic spin structure obtained from Rietveld refinements.

**Figure** 3 (a) Field dependent $\rho_{xx}$ measured above (upper panel) and below (lower panel) $T_N$. (b) Field dependent $\rho_{xy}$ measured at different temperatures. (c) Anomalous Hall resistivity $\Delta\rho_{xy}$ measured above $T_N$. (d) $\Delta\rho_{xy}$ (lower panel) and Hall conductivity $\sigma_{yx}$ (upper panel) measured below $T_N$. $\sigma_{yx}$ measured at $T = 20$ K is also shown as a reference.

**Figure** 4 (a) Color coded surface plot of anomalous Hall conductivity $\sigma^A$ as a function of temperature and applied field. (b) Anomalous Hall angle $\theta_{AHE}$ measured above (upper panel) and below (lower panel) $T_N$. (c) A comparison of the observed anomalous Hall response of DyPtBi with other reported systems. The grey curve is the data taken from field scan at $T = 4$ K ($T > T_N$)

and the blue curve is data taken from field scan at $T = 2$ K ($T < T_N$). Strong enhancement of the anomalous Hall response is evident.

## Acknowledgements

H. Z. is supported by the U.S. National Science Foundation under DMR-1608752. X. K. acknowledges the financial support by the U.S. Department of Energy, Office of Science, Office of Basic Energy Sciences, Materials Sciences and Engineering Division under DE-SC0019259. The work at Penn state is supported by the US National Science Foundation under grants DMR 1917579. Y.L.Z acknowledges partial financial support from the National Science Foundation through the Penn State 2D Crystal Consortium-Materials Innovation Platform (2DCC-MIP) under NSF cooperative agreement DMR-1539916. Access to MACS was provided by the Center for High Resolution Neutron Scattering, a partnership between the National Institute of Standards and Technology and the National Science Foundation under Agreement No. DMR-1508249. The identification of any commercial product or trade name does not imply endorsement or recommendation by the National Institute of Standards and Technology. A portion of this research used resources at the High Flux Isotope Reactor, a DOE Office of Science User Facility operated by Oak Ridge National Laboratory.

Figure 1.

Zhang et al,

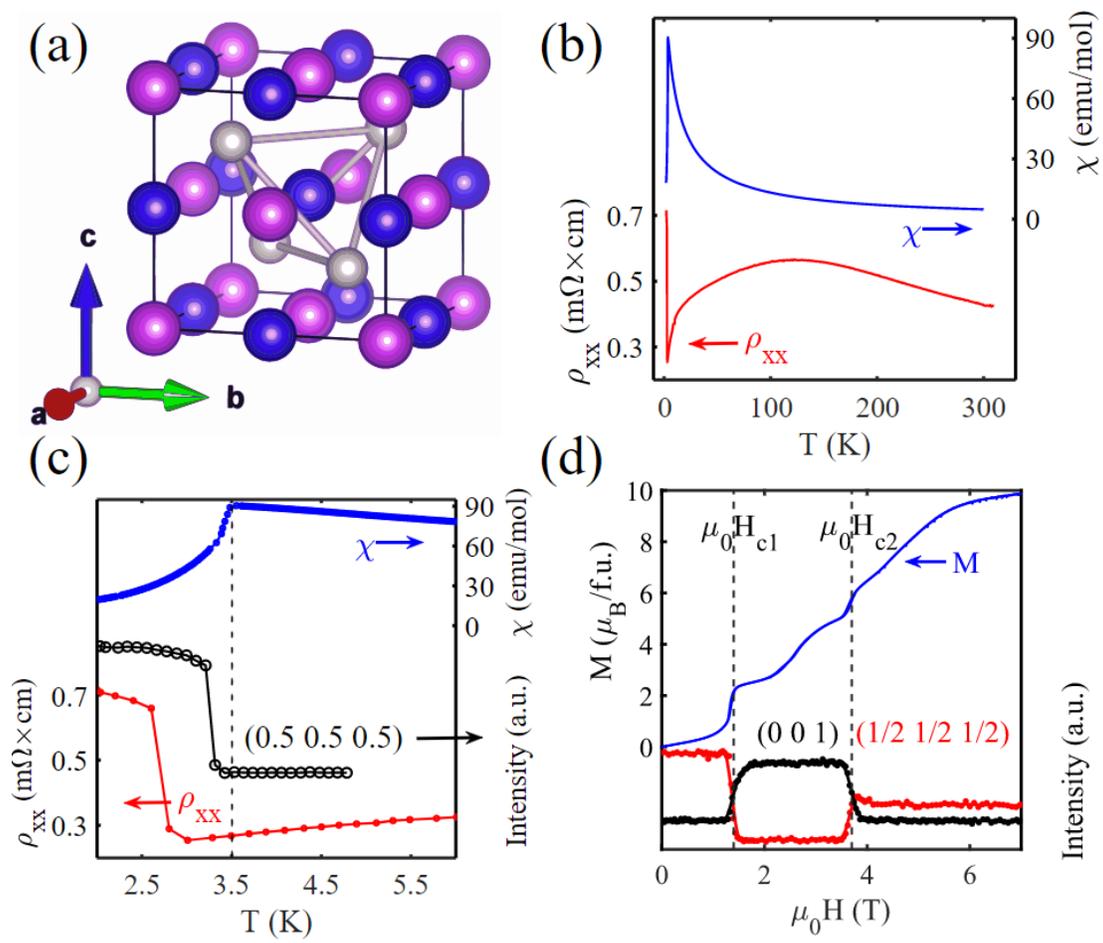

Figure 2.

Zhang et al,

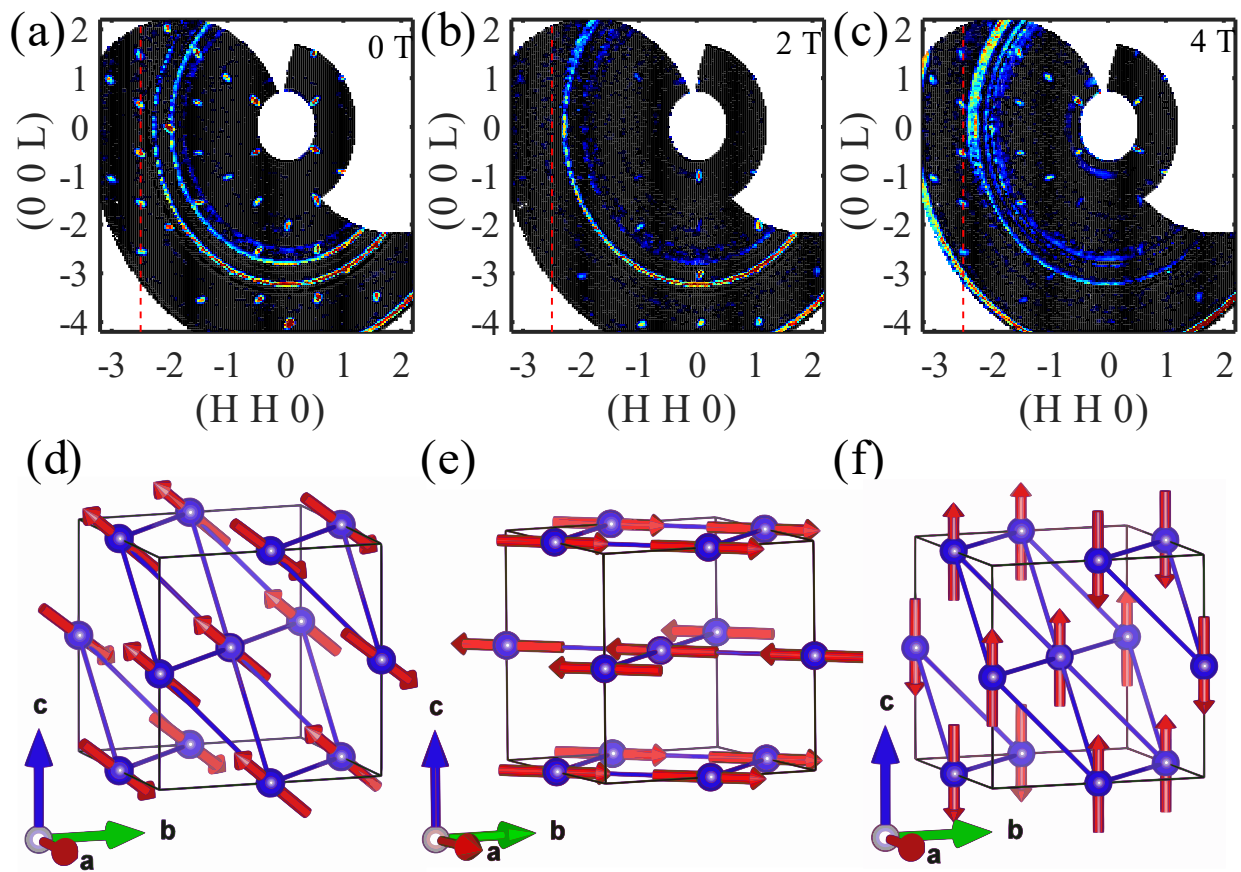

Figure 3.

Zhang et al,

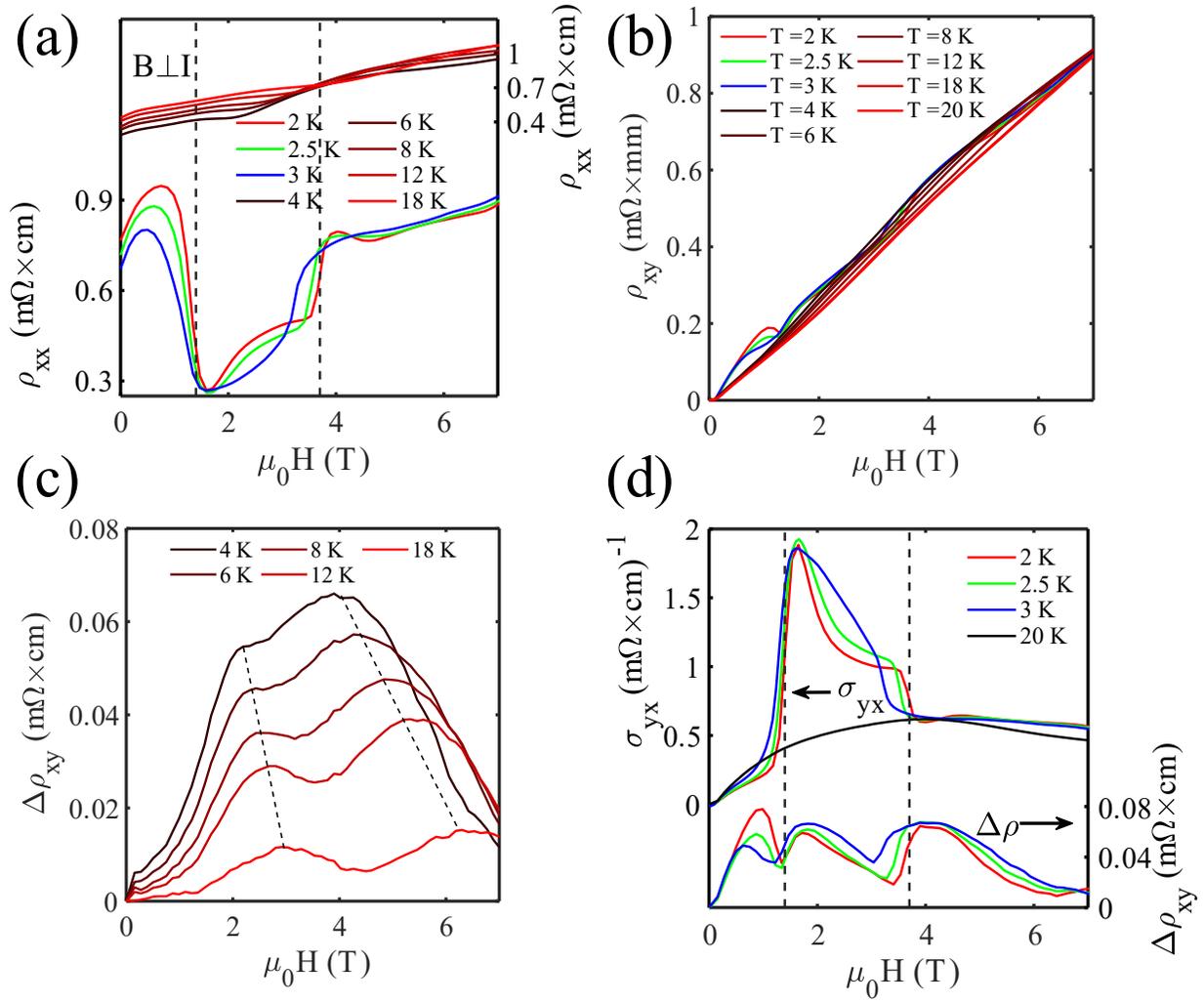

Figure 4.

Zhang et al,

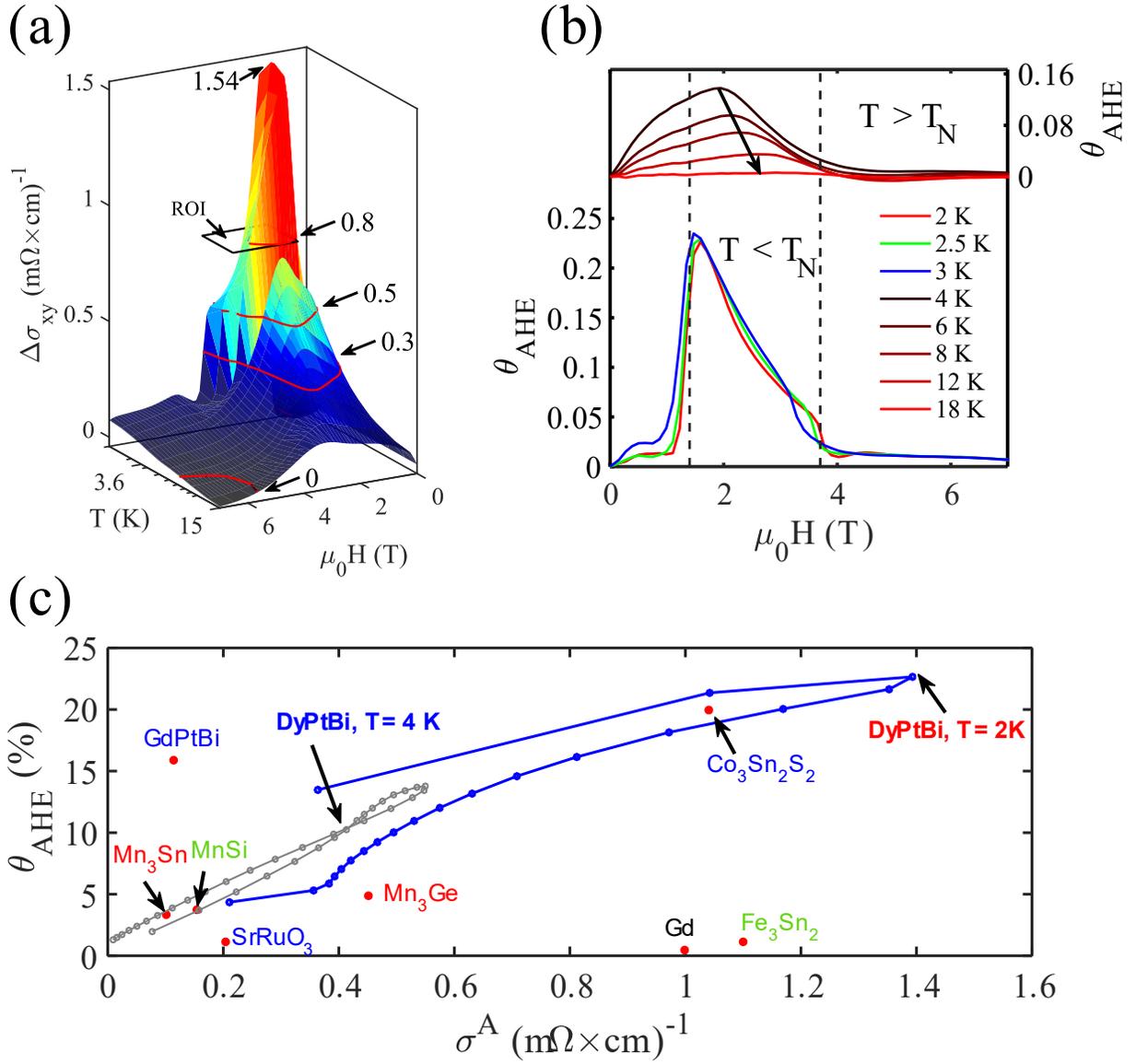

# References

[1] R. Li, J. Wang, X.-L. Qi, and S.-C. Zhang, Nature Physics **6**, 284 (2010).
[2] R. Yu, W. Zhang, H.-J. Zhang, S.-C. Zhang, X. Dai, and Z. Fang, Science **329**, 61 (2010).
[3] R. S. K. Mong, A. M. Essin, and J. E. Moore, Physical Review B **81**, 245209 (2010).
[4] X. Wan, A. M. Turner, A. Vishwanath, and S. Y. Savrasov, Physical Review B **83**, 205101 (2011).
[5] M. M. Otrokov, I. I. Klimovskikh, H. Bentmann, D. Estyunin, A. Zeugner, Z. S. Aliev, S. Gaß, A. U. B. Wolter, A. V. Koroleva, A. M. Shikin *et al.*, Nature **576**, 416 (2019).
[6] E. Liu, Y. Sun, N. Kumar, L. Muechler, A. Sun, L. Jiao, S.-Y. Yang, D. Liu, A. Liang, Q. Xu *et al.*, Nature Physics **14**, 1125 (2018).
[7] D. F. Liu, A. J. Liang, E. K. Liu, Q. N. Xu, Y. W. Li, C. Chen, D. Pei, W. J. Shi, S. K. Mo, P. Dudin *et al.*, Science **365**, 1282 (2019).
[8] N. Morali, R. Batabyal, P. K. Nag, E. Liu, Q. Xu, Y. Sun, B. Yan, C. Felser, N. Avraham, and H. Beidenkopf, Science **365**, 1286 (2019).
[9] I. Belopolski, K. Manna, D. S. Sanchez, G. Chang, B. Ernst, J. Yin, S. S. Zhang, T. Cochran, N. Shumiya, H. Zheng *et al.*, Science **365**, 1278 (2019).
[10] R. A. Müller, A. Desilets-Benoit, N. Gauthier, L. Lapointe, A. D. Bianchi, T. Maris, R. Zahn, R. Beyer, E. Green, J. Wosnitza *et al.*, Physical Review B **92**, 184432 (2015).
[11] T. Suzuki, R. Chisnell, A. Devarakonda, Y. T. Liu, W. Feng, D. Xiao, J. W. Lynn, and J. G. Checkelsky, Nature Physics **12**, 1119 (2016).
[12] Y. Zhu, B. Singh, Y. Wang, C.-Y. Huang, W.-C. Chiu, B. Wang, D. Graf, Y. Zhang, H. Lin, J. Sun *et al.*, Physical Review B **101**, 161105 (2020).
[13] O. Pavlosiuk, X. Fabreges, A. Gukasov, M. Meven, D. Kaczorowski, and P. Wiśniewski, Physica B: Condensed Matter **536**, 56 (2018).
[14] S. Chadov, X. Qi, J. Kübler, G. H. Fecher, C. Felser, and S. C. Zhang, Nature Materials **9**, 541 (2010).
[15] P. C. Canfield, J. D. Thompson, W. P. Beyermann, A. Lacerda, M. F. Hundley, E. Peterson, Z. Fisk, and H. R. Ott, Journal of Applied Physics **70**, 5800 (1991).
[16] P. C. Canfield and Z. Fisk, Philosophical Magazine B **65**, 1117 (1992).
[17] J. A. Rodriguez, D. M. Adler, P. C. Brand, C. Broholm, J. C. Cook, C. Brocker, R. Hammond, Z. Huang, P. Hundertmark, J. W. Lynn *et al.*, Measurement Science and Technology **19**, 034023 (2008).
[18] H. Cao, B. C. Chakoumakos, K. M. Andrews, Y. Wu, R. A. Riedel, J. Hodges, W. Zhou, R. Gregory, B. Haberl, and J. J. C. Molaison, Crystals **9**, 5 (2019).
[19] F. Keffer and H. Chow, Physical Review Letters **31**, 1061 (1973).
[20] Supplemental.
[21] J. Rodríguez-Carvajal, Physica B: Condensed Matter **192**, 55 (1993).
[22] T. Yildirim, A. B. Harris, and E. F. Shender, Physical Review B **58**, 3144 (1998).
[23] B. G. Ueland, A. Kreyssig, K. Prokeš, J. W. Lynn, L. W. Harriger, D. K. Pratt, D. K. Singh, T. W. Heitmann, S. Sauerbrei, S. M. Saunders *et al.*, Physical Review B **89**, 180403 (2014).
[24] J. Wosnitza, G. Goll, A. D. Bianchi, B. Bergk, N. Kozlova, I. Opahle, S. Elgazzar, M. Richter, O. Stockert, H. v. Löhneysen *et al.*, New Journal of Physics **8**, 174 (2006).
[25] C. Liu, Y. Lee, T. Kondo, E. D. Mun, M. Caudle, B. N. Harmon, S. L. Bud'ko, P. C. Canfield, and A. Kaminski, Physical Review B **83**, 205133 (2011).
[26] N. Nagaosa, J. Sinova, S. Onoda, A. H. MacDonald, and N. P. Ong, Reviews of Modern Physics **82**, 1539 (2010).
[27] Z. Fang, N. Nagaosa, K. S. Takahashi, A. Asamitsu, R. Mathieu, T. Ogasawara, H. Yamada, M. Kawasaki, Y. Tokura, and K. Terakura, Science **302**, 92 (2003).
[28] T. Miyasato, N. Abe, T. Fujii, A. Asamitsu, S. Onoda, Y. Onose, N. Nagaosa, and Y. Tokura, Physical Review Letters **99**, 086602 (2007).

[29] L. Ye, M. Kang, J. Liu, F. von Cube, C. R. Wicker, T. Suzuki, C. Jozwiak, A. Bostwick, E. Rotenberg, D. C. Bell *et al.*, Nature **555**, 638 (2018).
[30] A. K. Nayak, J. E. Fischer, Y. Sun, B. Yan, J. Karel, A. C. Komarek, C. Shekhar, N. Kumar, W. Schnelle, J. Kübler *et al.*, Science Advances **2**, e1501870 (2016).
[31] S. Nakatsuji, N. Kiyohara, and T. Higo, Nature **527**, 212 (2015).
[32] N. Manyala, Y. Sidis, J. F. DiTusa, G. Aeppli, D. P. Young, and Z. Fisk, Nature Materials **3**, 255 (2004).
[33] R. Shindou and N. Nagaosa, Physical Review Letters **87**, 116801 (2001).
[34] D. Grohol, K. Matan, J.-H. Cho, S.-H. Lee, J. W. Lynn, D. G. Nocera, and Y. S. Lee, Nature Materials **4**, 323 (2005).